\documentclass[12pt]{iopart}

\usepackage{etoolbox}

\makeatletter
\newcommand{\mainmatter}{%
  \setcounter{footnote}{0}%
  \patchcmd{\@makefntext}{\fnsymbol}{\arabic}{}{}%
  \patchcmd{\@thefnmark}{\fnsymbol}{\arabic}{}{}%
  \def\@makefnmark{\textsuperscript{\arabic{footnote}}}%
}
\makeatother

\expandafter\let\csname equation*\endcsname\relax
\expandafter\let\csname endequation*\endcsname\relax
\usepackage{amsmath,amssymb,amsfonts}
\usepackage{bm}
\usepackage{color}
\usepackage[normalem]{ulem} 

\usepackage{graphicx}
\usepackage{indentfirst}
\usepackage{psfrag}
\usepackage{epsfig}
\usepackage{color} 

\newcommand{\del}{\partial}

\newcommand{\QPEa}{\ensuremath{\mathcal{Q}_1^{\textrm{(PE)}}}}
\newcommand{\QPEatilde}{\ensuremath{\tilde{\mathcal{Q}}_1^{\textrm{(PE)}}}}

\newcommand{\QMB}{\ensuremath{\mathcal{Q}_1^{\textrm{(MB)}}}}
\newcommand{\neff}{\ensuremath{n_\text{eff}}}
\newcommand{\Wm}{\ensuremath{(\text{Wm})^{-1}}}

\newcommand{\kscore}{\ensuremath{k_\text{s}^{(1)}}}
\newcommand{\kpcore}{\ensuremath{k_\text{p}^{(1)}}}
\newcommand{\ksclad}{\ensuremath{k_\text{s}^{(2)}}}
\newcommand{\kpclad}{\ensuremath{k_\text{p}^{(2)}}}
\newcommand{\kscorezero}{\ensuremath{k_{\text{s},0}^{(1)}}}

\newcommand{\kscladzero}{\ensuremath{k_{\text{s},0}^{(2)}}}

\newcommand{\ksj}{\ensuremath{k_\text{s}^{(j)}}}

\newcommand{\klj}{\ensuremath{k_\text{p}^{(j)}}}

\newcommand{\revision}[1]{{#1}}

\begin{document}

	\title{ARRAW: Anti-resonant reflecting acoustic waveguides}
 	\author{M K Schmidt$^{1}$\footnote{Corresponding author.}\footnote{These two authors contributed equally.}, 
 	    M C O'Brien$^{1,2}\S$, 
 	    M J Steel$^{1}$ 
 	    and C G Poulton$^{3}$}
	\address{$^1$ Macquarie University Research Centre in  Quantum Engineering (MQCQE), MQ Photonics Research Centre, Department of Physics and Astronomy, Macquarie University, NSW 2109, Australia.} 
    \address{$^2$ School of Physics, The University of New South Wales, NSW 2052, Australia.}
    \address{$^3$ School of Mathematical and Physical Sciences, University of Technology Sydney, NSW 2007, Australia.}	\eads{\mailto{mikolaj.schmidt@mq.edu.au}, \mailto{ matthew.obrien@unsw.edu.au}, \mailto{michael.steel@mq.edu.au}, \mailto{christopher.poulton@uts.edu.au}}
	
	\begin{abstract}
		Development of acoustic and optoacoustic on-chip technologies calls for new solutions to guiding, storing and interfacing acoustic and optical waves in integrated silicon-on-insulator (SOI) systems. One of the biggest challenges in this field is to suppress the radiative dissipation of the propagating acoustic waves, while co-localizing the optical and acoustic fields in the same region of an integrated waveguide. Here we address this problem by introducing Anti-Resonant Reflecting Acoustic Waveguides (ARRAWs) --- mechanical analogues of the Anti-Resonant Reflecting Optical Waveguides (ARROWs). We discuss the principles of anti-resonant guidance and establish guidelines for designing efficient ARRAWs. Finally, we demonstrate examples of the simplest silicon/silica ARRAW platforms that can simultaneously serve as near-IR optical waveguides, and support strong backward Brillouin scattering.
	\end{abstract}		
	\maketitle

\section{Introduction}
Almost all conventional optical step-index waveguides are unsuitable for confining and supporting the low-loss propagation of acoustic waves. This is because the high refractive index materials making up the core of optical waveguides tend to support acoustic waves propagating at larger velocities than in the low-refractive index cladding layers \cite{eggleton2019brillouin}. Consequently, acoustic waves do not experience total internal reflection (TIR) at the core-cladding interface, and dissipate by free propagation into the cladding. Conversely, acoustic waveguides relying on a reversed design --- with the acoustically \textit{slow} material making up the core, and the \textit{fast} material the cladding --- usually do not guide optical waves,
due to a general association between refractive index and material density. Therefore, if we are to pursue systems implementing efficient interaction between propagating optical and acoustic waves, and particularly Brillouin scattering, we need to look beyond the simple physics of TIR.

A number of designs have been put forward to address this challenge \cite{eggleton2019brillouin,Safavi-Naeini:19,aspelmeyer2014cavity,eggleton2013inducing}. In some, the waveguides are suspended in air by either sparsely positioned \cite{shin2013tailorable,kittlaus2016large,van2015net} or specifically engineered supporting structures \cite{schmidt19}. In others, both light and sound are guided along line defects of phoxonic crystals \cite{doi:10.1063/1.2216885,zhang2017design,Yu:18}. Finally, a combination of the desired material properties --- high refractive index and low stiffness --- has been identified in chalcogenides, allowing researchers to revisit step-index architectures for optoacoustic waveguides \cite{Pant:11, morrison2017compact}.

In this work, we suggest a simple, novel class of waveguides capable of supporting the simultaneous propagation of co-localized optical and acoustic waves, based on the concept of Anti-Resonant Reflection Optical Waveguides (ARROWs) \cite{Litchinitser:03}. ARROWs were originally studied to enable low-loss optical guidance in the earliest integrated optical waveguides \cite{duguay1986antiresonant}. At that time, integrated photonic devices relied on a small contrast of refractive index inducing TIR between the doped silica medium making up the core, and the pure silica of the cladding. In ARROWs however, this design was inverted, allowing light to be guided in a low-refractive-index (fast) core, surrounded by a high-refractive-index ({slow}) cladding. This is achieved by engineering the cladding to behave like a Fabry-Perot layer operating at the anti-resonance condition \cite{archambault,Litchinitser:02}. Variations of ARROWs are now widely used in liquid core waveguides developed for biomolecular detection \cite{C0LC00496K,testa2016liquid,7282086}. 

\begin{figure}[htbp!]
	\begin{center}
		\includegraphics[width=.5\columnwidth]{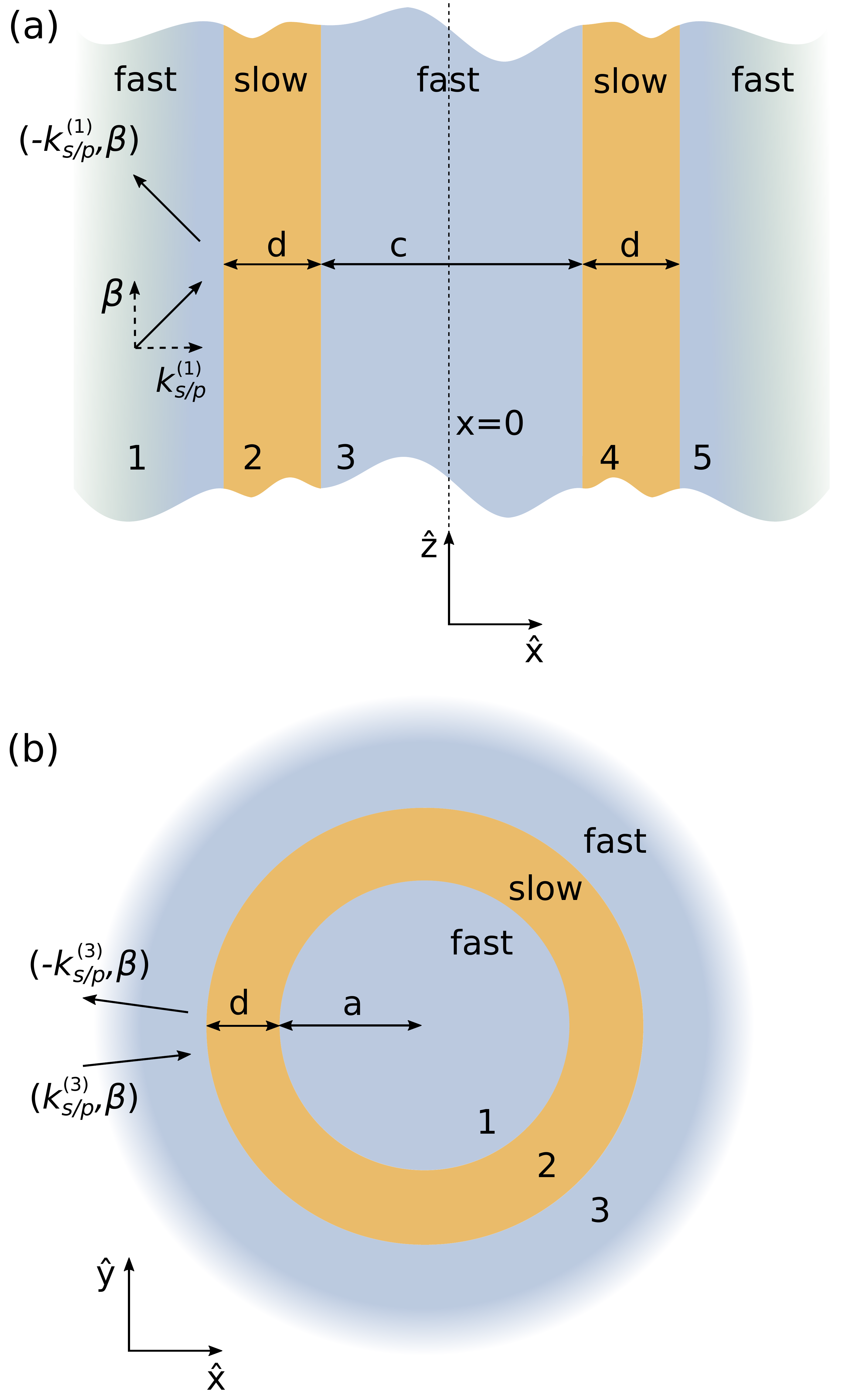}
		\caption{Schematics of (a) planar and (b) cylindrical waveguides. In both systems the acoustically \textit{fast} core is made of the same material as the outermost cladding, and supports both longitudinal ($P$) and shear ($S$) waves with velocities larger than the respective velocities in the \textit{slow} {inner claddings}. The propagation along the waveguide ($\hat{\mathbf{z}}$ axis) is characterized by the wavenumber $\beta$ which is constant throughout the structure, and the transverse wavenumbers $k_{s/p}^{(j)}$ in medium `j'.}
		\label{Fig1}
	\end{center}
\end{figure}

As we show here, the acoustic analogue of such waveguides --- Anti-Resonant Reflecting Acoustic Waveguides (ARRAWs), are capable of guiding acoustic waves through an acoustically fast core due to anti-resonances in the acoustically slow cladding. For example, in the particular designs of silicon/silica/silicon planar and cylindrical waveguides depicted in figure~\ref{Fig1}, the acoustic field of ARRAW modes would be predominantly localized to the silicon core. 

Furthermore, such ARRAWs can simultaneously support the conventional TIR guidance of light in the high-refractive index core, and consequently amplify local optoacoustic interactions between the co-localized optical and acoustic waves, including Brillouin scattering. Out of the two interaction mechanisms previously identified as contributing to Brillouin effects: photoelasticity and radiation pressure \cite{rakichPRX,sipe2016hamiltonian,wolff2015stimulated}, the former relies on the acoustic field locally modifying the refractive index of the bulk of the medium, forming a moving grating for the optical fields. This effect necessarily relies on the spatial overlap between the optical and acoustic field, and the built-up amplitude of the induced acoustic field, quantified by the mechanical quality factor of the acoustic mode $Q_m$. 
Here we will analyze in detail how ARRAWs can be optimized to support high-$Q_m$ acoustic, as well as optical modes, co-localized in the core of the waveguide, giving rise to strong backwards Brillouin scattering.

The paper is structured as follows: In the two following sections we discuss ARRAW behavior in planar and cylindrical waveguides. Since sound can propagate in a solid medium in the form of both transverse waves (referred to throughout as \textit{S} waves) and longitudinal waves (\textit{P} waves), characterized by different velocities, we expect that ARRAW waveguides will exhibit a richer structure of modes than their optical counterparts.  For each structure we highlight special, optics-like cases in which shear waves become decoupled from the longitudinal components, and the system is an analogue of an anti-resonant \textit{optical} waveguide. Finally, we demonstrate how cylindrical ARRAW waveguides can be used to support simultaneous and co-localized optical and acoustic modes in the core of the waveguide, and discuss how such structures thus support efficient backward Brillouin scattering. 

\section{ARRAW condition in planar structures}

In this section, we consider the simplest geometry of a planar waveguide, shown schematically in figure~\ref{Fig1}(a). The central section of the waveguide --- the \textit{core} (marked as `3') --- is made up of the same material as the semi-infinite {outermost cladding layers} (`1' and `5'), and supports \textit{S} waves with velocity $v_\text{s}^{(1)}$, larger than the respective velocity $v_\text{s}^{(2)}$ in the {inner cladding layers} `2' and `4' ($v_\text{s}^{(1)}>v_\text{s}^{(2)}$). Similar ordering is established for the velocities of \textit{P} waves in the core and cladding ($v_\text{p}^{(1)}>v_\text{p}^{(2)}$). While a more realistic design of the waveguide would include a finite outermost cladding layer surrounded by air, we focus on the semi-infinite model to stress that the acoustic guidance is induced solely by the anti-resonance in the cladding layer, rather than reflection from the solid-air interface. \revision{We also make two approximations about all the considered materials, neglecting their viscosities, and assuming isotropic elastic responses (see \ref{AppendixC} for the values of elastic parameters used here). These approximations allow us to develop tractable analytical models for the physics of ARRAW. Furthermore, for the application in Brillouin scattering \cite{eggleton2019brillouin}, we will only consider modes with larger quality factors $Q_m>100$, for which the small losses introduced by viscosity can be considered as additive to the radiative dissipation \cite{rakichPRX}}. Here we demonstrate the ARRAW behavior in the particular platform consisting of a silicon waveguide core and outer cladding, and a silica inner cladding. This choice is motivated by the advancement of the fabrication protocols for these materials, and the significant interest in implementing platforms for Brillouin interaction in silicon \cite{eggleton2019brillouin}. 

The waveguiding modes of these structures are found by solving the elastic equation of motion in each layer $j$,
\begin{equation}\label{wave.equation}
	\rho^{(j)} \frac{\textrm{d}^2}{\textrm{d}t^2}\mathbf{u}^{(j)} = \nabla \cdot \mathbf{T}^{(j)},
\end{equation}
relating the density $\rho^{(j)}$, displacement field $\mathbf{u}^{(j)}$, and stress tensor $\mathbf{T}^{(j)}$ \cite{auld1973acoustic}. Stress is related to strain $\mathbf{S}^{(j)}=\nabla_S\mathbf{u}^{(j)}$ --- the symmetrized gradient of the displacement field --- via Hooke's law and the rank 4 stiffness tensor $\mathbf{c}^{(j)}$, by the dyadic product $\mathbf{T}^{(j)}=\mathbf{c}^{(j)}:\mathbf{S}^{(j)}$. In isotropic media, this relationship can be expressed through the Lam\'e parameters $\lambda^{(j)}$ and $\mu^{(j)}$ as $T^{(j)}_{kl} = 2\mu^{(j)}{S}^{(j)}_{kl}+\lambda^{(j)} \delta_{kl}{S}^{(j)}_{nn}$. To obtain the solution, we use the ansatz
\begin{equation}\label{harmonicsolution}
	\mathbf{u}(\mathbf{r}_{\perp},z,t) = \mathbf{U}(\mathbf{r}_{\perp}) \rme^{\rmi(\beta z - \Omega t)} + \text{c.c.},
\end{equation}
where c.c. denotes the complex conjugate, $\Omega$ and $\beta$ are the angular frequency and the longitudinal wavenumber of the mode, respectively, and $\mathbf{r}_{\perp}$ is the transverse coordinate. The fields in neighboring layers are related via the elastic boundary conditions, which require the continuity of the normal components of the adjacent stress tensors, and all the components of the adjacent displacement fields \cite{auld1973acoustic}. The waveguide modes are found by requiring that the amplitudes of incoming fields in the outer-most layers (`1' and `5' in the planar structure --- see figure~\ref{Fig1}(a)) vanish. For a planar structure, these modes are found in the basis of shear and longitudinal planewaves in each of the layers, with the transverse wavevectors in medium `j' denoted by \ksj and \klj, respectively. Details of the calculations are given in \ref{AppendixA}.

\subsection{Out-of-plane  polarization (pure shear)}

We first consider the case of \textit{pure shear waves}, with solely out-of-plane displacement fields ($u^{(j)}_x=u^{(j)}_z=0$). These are uncoupled from longitudinal modes throughout the structure (see the derivations in \ref{AppendixA} or \cite{auld1973acoustic}). The guidance condition (vanishing amplitude of incoming waves in the outermost layer) yields a rather complex transcendental equation relating $\Omega$, $\beta$ and $k_{s/p}^{(j)}$ (not shown here). However, it can be simplified by considering separately the modes which are symmetric and anti-symmetric with respect to the $\hat{\mathbf{y}}$-$\hat{\mathbf{z}}$ symmetry plane of the structure at $x=0$. The resulting transcendental equations for the symmetric and anti-symmetric modes can be expressed as
\begin{equation}\label{sym}
\left(-1+\rme^{\rmi c \kscore}r_s\right)-\rme^{2 \rmi d \ksclad}r_s\left(-r_s+\rme^{\rmi c \kscore}\right)=0,
\end{equation}
\begin{equation}\label{asym}
\left(1+\rme^{\rmi c \kscore}r_s\right)-\rme^{2 \rmi d \ksclad}r_s\left(r_s+\rme^{\rmi c \kscore}\right)=0,
\end{equation}
respectively, where as marked in figure~\ref{Fig1}, $c$ and $d$ are the thicknesses of the core and cladding layers, and
\begin{equation}
r_s = \frac{\kscore \mu^{(1)}-\ksclad \mu^{(2)}}{\kscore \mu^{(1)}+\ksclad \mu^{(2)}},
\end{equation}
is the acoustic reflection coefficient for a pure \textit{S} wave propagating in medium `1', reflecting off the interface with medium `2'. 

\subsubsection{Optics-like anti-resonance condition}

We note that equations \eqref{sym} and \eqref{asym} also describe the symmetric and antisymmetric \textit{s}-polarized \textit{optical} waveguiding modes, if we replace $r_s$ with the optical reflection coefficient for non-magnetic materials $r^\text{opt} = (k^{(1)}-k^{(2)})/(k^{(1)}+k^{(2)})$. This is thanks to the purely transverse nature of these modes, and the close mapping between acoustic and optical boundary conditions (with the continuity of tangential components of the electric field, and of normal components of the magnetic field, mirroring the continuity of acoustic displacement and normal stress components, respectively).

Furthermore, to further explore the analogy to optics for ARRAWs, we can simplify the conditions given in equations \eqref{sym} and \eqref{asym} by eliminating the dependence on the thickness of the core layer ($c$) from the transcendental equations. To this end, we choose the core radius $c$ to correspond to the lowest order \textit{symmetric} modes supported by the core:\cite{Litchinitser:03} 
\begin{equation}\label{sym.2}
\kscore c = (2n+1)\pi,
\end{equation} 
for $n=0,1,2,\dots$, and arrive at the simplified symmetric ARRAW condition from \eqref{sym}
\begin{equation}\label{condition.core}
	\rme^{2 \rmi \ksclad d} = r_s.
\end{equation}
Similarly, the lowest order \textit{antisymmetric} modes \eqref{asym} are found for a core width satisfying
\begin{equation}\label{asym.2}
 \kscore c = 2n \pi, 
\end{equation}
which simplifies \eqref{asym} to the ARRAW condition given in \eqref{condition.core}, identical to the symmetric ARRAW modes.

To simplify this condition even further, we can consider waves propagating in the core at a glancing incidence to the cladding interface, where the reflection coefficient $r_s\approx -1$. We then retrieve from \eqref{condition.core} the approximate relation:
\begin{equation}\label{arraw}
	\ksclad d = (2m+1)\frac{\pi}{2},
\end{equation} 
for $m=0,1,2,\dots$, which can be used to identify anti-resonant behavior in the dispersion of acoustic waveguides. This condition is analogous to that found for ARROWs, which reads $k^{(2)} d = (2m+1){\pi}/{2}$.


\subsubsection{Exact dispersion relation}

\begin{figure*}[htbp!]
	\includegraphics[width=.9\textwidth]{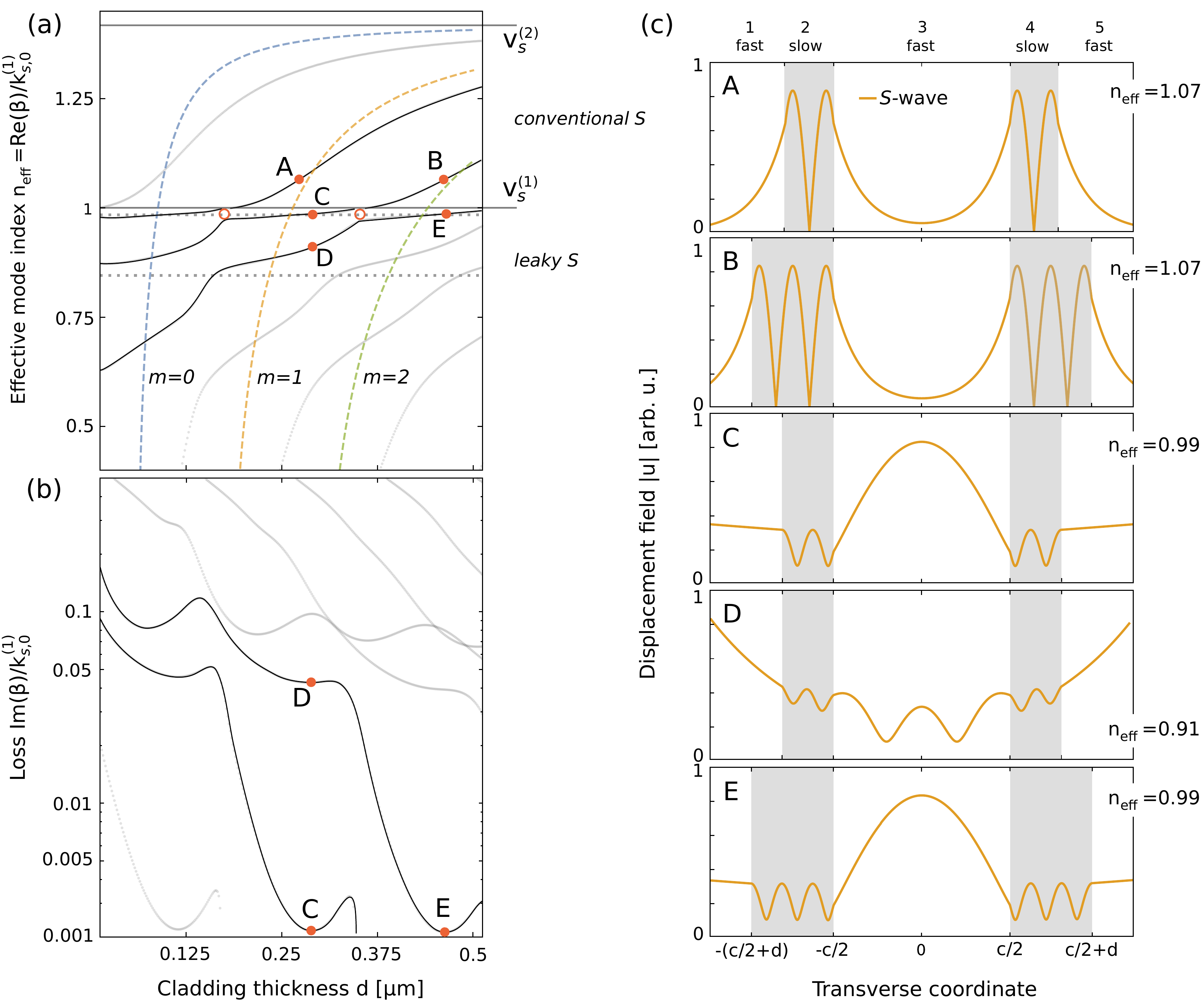}
	\caption{(a) Dispersion relation, (b) normalized loss, and (c) field distribution plots for out-of-plane polarized symmetric shear modes of a planar waveguiding structure \revision{with core width $c=1~\mu$m}. In (a) horizontal gray dotted lines correspond to the analytic conditions for odd core resonance \eqref{sym.2} with $n=0,~1$, and colored dashed lines are described by the cladding anti-resonance condition given in \eqref{arraw} with $m=0,1$ and $2$. Thicker black lines denote the dispersion of conventionally guided (A-B) and leaky shear (C-E) modes shown in (c).}	
	\label{Fig2}
\end{figure*}

The complete, exact dispersion relation of the symmetric, pure shear modes of the planar waveguide with arbitrary core radius $c$ is found by solving \eqref{sym} numerically, and includes families of both conventionally guided and leaky modes. To differentiate between the two, and to establish a parallel between acoustic and optical frameworks, we arbitrarily choose the largest shear velocity of the system $v_\text{s}^{(1)}$ as a reference, and define an acoustic \textit{effective mode index}
\begin{equation}
    n_{\text{eff}} = \frac{\mathrm{Re}(\beta)}{\kscorezero},
\end{equation}
where $\kscorezero=\Omega/v_\text{s}^{(1)}$. Conventionally guided modes, characterized by a real longitudinal wavenumber $\beta$, and the acoustic field propagating predominantly in the slow cladding layer thus correspond to $1 < \neff < v_\text{s}^{(1)}/v_\text{s}^{(2)}$. Conversely, leaky modes found for $n_{\text{eff}} < 1$ have complex longitudinal wavenumber.

These features are demonstrated in figure~\ref{Fig2}, where we analyze the symmetric, pure shear modes of a Si(core)/SiO$_{2}$(inner cladding)/Si(outer cladding) planar waveguide. For this setup, the ratio of shear velocities $v_\text{s}^{(1)}/v_\text{s}^{(2)}\approx 1.42$. The core width is set to $c=1~\mu$m, and we vary the cladding width $d$ from approximately 0 to $d=0.5~\mu$m --- a range of sizes which, for acoustic modes propagating at frequency $\Omega/2\pi= 15$~GHz, spans $\kscladzero d\approx 0$ to $4\pi$. In figure~\ref{Fig2}(a) we plot the effective mode index $n_{\text{eff}}$, and in (b) the normalized attenuation parameter $\mathrm{Im}(\beta)/\kscorezero$. 

In figure~\ref{Fig2}(a) we clearly identify the families of modes conventionally guiding the acoustic waves in the inner cladding layer, characterized by $1<\neff<v_\text{s}^{(1)}/v_\text{s}^{(2)}$. These modes exhibit purely real transverse wavenumbers in the cladding (\ksclad), and purely imaginary transverse wavenumbers in the core (\kscore). This indicates oscillatory behavior of the fields in the cladding, along the transverse direction $\hat{\mathbf{x}}$, and exponential localization of the fields to the interfaces in the core. We clearly identify these features in the two upper panels of figure~\ref{Fig2}(c), which show the displacement field profile of two modes depicted as A and B in figure~\ref{Fig2}(a,b), differentiated by the number of nodes of the displacement field in the cladding (with $\ksclad d \approx 3\pi/2$ in A and $5\pi/2$ in B).


Modes below the shear core sound line ($\neff<1$) are characterized by complex wavenumbers $\beta$, and $\ksj$'s, meaning that the field in the core does not simply exponentially decay with $x$ away from the core/cladding interface, but exhibits oscillations governed by $\mathrm{Re}(\kscore)$. Simultaneously, in a manner consistent with optical \textit{leaky} modes, the imaginary part of $\kscore$ becomes negative, and the outgoing fields increase exponentially with $x$ outside the structure. Just below the $\neff=1$ line, the glancing modes form flat-dispersion sections where acoustic waves are confined predominantly to the core (see panels C and E in figure~\ref{Fig2}(c); modes C, D and E were selected to match the local minima of {normalized loss} $\mathrm{Im}(\beta)/k_s^{(1}$). This behavior is described approximately by the horizontal gray dotted lines depicting resonances of the core ($n=0$ in \eqref{sym.2}), which cross with the colored dashed lines indicating the cladding anti-resonance ($m=0,1,2$ in \eqref{arraw}) near the minimum of the loss function. This agreement breaks down for mode D, which lies far below the $\neff=1$ line, and does not meet the glancing incidence criterion. We also identify a prominent anti-crossing behavior near the crossings between the resonances of the core (horizontal dotted gray lines) and resonances of the cladding ($\ksclad d= m\pi$, anti-crossings marked with hollow circles). At these points the cladding transmission reaches a local maximum, suppressing the formation of a waveguiding mode.

This simple analysis allows us to identify the ARRAW modes as the leaky acoustic modes localized to the \textit{fast} medium, and found at the local minima of loss, right below the \textit{sound-line} corresponding to the velocity of waves in the fastest medium (core) --- here illustrated as modes C and E. As we show below, this definition naturally extends to polarizations and media supporting the propagation of \textit{P} waves.


\subsection{In-plane polarization}

In-plane polarization of the waveguiding modes ($u_y^{(i)}=0$) necessarily couples the in-plane shear (\textit{S}) and longitudinal (\textit{P}) waves. 
Consequently, the transcendental equation for the modes is more complex (see derivation in \ref{AppendixA}), even if we focus on a particular symmetry of both the components. Nevertheless, it submits to numerical solution. As in the out-of-plane polarization case, we focus on symmetric modes for simplicity. Furthermore, to observe anti-resonant behavior of longitudinal waves, characterized by substantially longer wavelengths (since $v_\text{p}^{(2)}/v_\text{s}^{(2)}\approx 1.6$), we consider claddings of larger thickness, comparable to the longitudinal wavelength in the cladding medium.

\begin{figure*}[htbp!]
	\includegraphics[width=.9\textwidth]{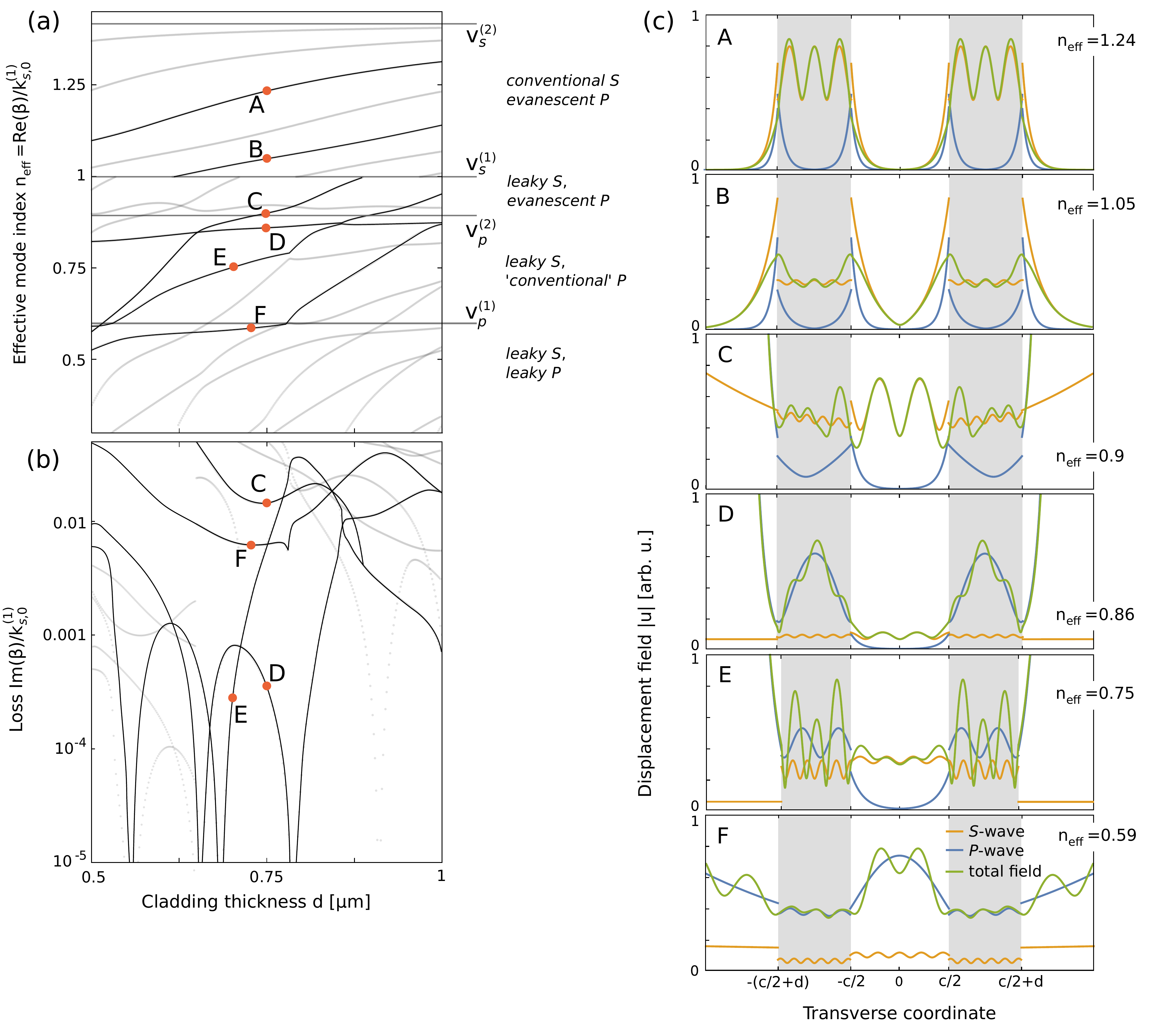}
	\caption{(a) Dispersion relation, (b) normalized loss, and (c) field distribution plots for in-plane polarization symmetric modes of a planar waveguiding structure \revision{with core width $c=1~\mu$m}, with orange and blue lines respectively denoting shear and  longitudinal contributions to the total (green lines) field, which is continuous at interfaces. Note that the thicknesses of cladding layers $d$ differ slightly for various modes. Characteristics of the selected modes are described in details in the text.}	
	\label{Fig3}
\end{figure*}

In the following discussion, it is useful to draw from the analytical framework presented in \ref{AppendixA}, and treat \textit{S} (shear) and \textit{P} (longitudinal) components of the acoustic wave (denoted as $\mathbf{u}_s$ and $\mathbf{u}_p$) as if they were independent, and characterize regimes in which these components behave as evanescent (exponentially decaying away from interfaces in either direction), conventionally guided (in the inner cladding) or leaky waves. While, as we show below, the coupling between the \textit{S} and \textit{P} waves necessarily blurs the characteristics of these regimes, this approach is instructive in developing intuition about the in-plane polarization acoustic guidance.

The dispersion relations and loss of the symmetric, in-plane modes are shown in figure~\ref{Fig3}(a) and (b), respectively. The field distributions (separated into \textit{S} and and \textit{P} components described as the norms of $\mathbf{u}_s$ and $\mathbf{u}_p$ fields defined in \ref{AppendixA}) of modes marked as A-F are shown in (c), chosen to best represent 4 distinct regimes of acoustic guidance. The first, denoted in figure~\ref{Fig3}(a) as \textit{conventional S} and 
\textit{evanescent P} ($1<\neff<v_\text{s}^{(1)}/v_\text{s}^{(2)}$), is characterized by real $\beta$, \ksclad, and imaginary \kscore (indicating conventional \textit{S} guidance in the cladding), as well as purely imaginary transverse \textit{P} wave $k_p^{(i)}$'s (indicating \textit{surface states} with fields exponentially decaying away from the interface in the core and the outer layer). Two examples of such modes are shown in panels A and B in figure~\ref{Fig3}(c). 

For $\neff < 1$, $\beta$ becomes complex, and the shear components of the fields diverge exponentially outside of the structure as in the out-of-plane case (see panels C-F), as expected for the leaky \textit{S} modes. In particular, as \neff~is reduced below $v_\text{s}^{(1)}/v_\text{p}^{(2)}\approx 0.89$, we first find the regime ($\neff>v_\text{s}^{(1)}/v_\text{p}^{(1)}\approx 0.6$) in which \textit{P} components become conventionally guided in the inner cladding (see modes D and E). Modes found in this regime exhibit very particular characteristics, as they mix the conventional-like localization of the \textit{P} waves with the leaky-like exponential increase of displacement fields in the outer cladding layer (see blue lines in D and E) due to the negative imaginary component of $\kpcore$ (resulting from complex $\beta$, or equivalently, coupling to the \textit{S} waves). Furthermore, for particular geometric parameters the normalized loss of these modes (figure~\ref{Fig3}(b)) appears to dip towards 0 --- a behavior which we identify with the onset of the simultaneous anti-resonant guidance of \textit{S} waves (note the localization of \textit{S} waves represented by orange lines to the core in D and E) and conventional \textit{P} guidance.



Finally, further reducing \neff~below $v_\text{s}^{(1)}/v_\text{p}^{(1)}$ brings us to the leaky \textit{S} and \textit{P} regime, 
in which the displacement fields of both the \textit{P} and \textit{S} contributions oscillate in the core and in the inner cladding layer. We can therefore expect to find modes for which both the \textit{P} and \textit{S} waves build up an anti-resonance in the inner cladding, by approximately meeting the ARRAW condition given for \textit{S} waves in \eqref{arraw}. These conditions will not be fulfilled exactly, since the transverse wavenumbers \ksclad and \kpclad are imaginary, and the two components are coupled. Nevertheless, we can find ARRAW modes, such as the one shown in panel F, which meet our previous definition: they are characterized by a local minimum of loss, lie immediately below the $\neff=v_\text{s}^{(1)}/v_\text{p}^{(1)}$ sound line, and exhibit a strong localization of both the \textit{P} and \textit{S} components in the core.


We should also note that the combination of Si/SiO$_2$ materials forbids the formation of modes with both \textit{S} and \textit{P} components which are conventionally guided, since the regions of effective mode velocities $(v_\text{s}^{(2)},v_\text{s}^{(1)})$ and $(v_\text{p}^{(2)},v_\text{p}^{(1)})$ do not overlap. Such regions could be found for other combinations of materials, e.g. Si/As$_2$S$_3$ or SiO$_2$/As$_2$S$_3$.

\section{ARRAW modes in cylindrical waveguides}

In order to bring the concept of ARRAW closer to applications in nonlinear optical system, we now investigate ARRAW behavior in cylindrical waveguides, as shown schematically in figure~\ref{Fig1}(b). In these designs, the core (medium `1') of radius $a$ is surrounded by a cladding `2' of thickness $d$ and, as for layered waveguides, the core and outer semi-infinite layer `3' are made up of the same material. The mathematical formulation of this problem is discussed in detail in \ref{AppendixB}, where we expand the fields into a basis of torsional (\textit{S} wave components only) and dilatational modes. For simplicity, we will consider only the azimuthally symmetric case ($m=0$), in which the two are decoupled, and the torsional modes have azimuthal component $\mathbf{u} = u_\theta\hat{\bm{\theta}}$ only \cite{auld1973acoustic} (see \ref{AppendixB}). We thus arrive at a system much like that found in planar waveguides, where one family of modes (torsional) depends on the shear velocities only, while the other family (dilatational) mixes \textit{S} and \textit{P} components. We thus expect to recover the optics-like ARROW characteristics of the former, and the ARRAW-like, complex modes of the latter.

\subsection{Torsional modes}
\label{section31}

\begin{figure*}[htbp!]
	\includegraphics[width=.9\textwidth]{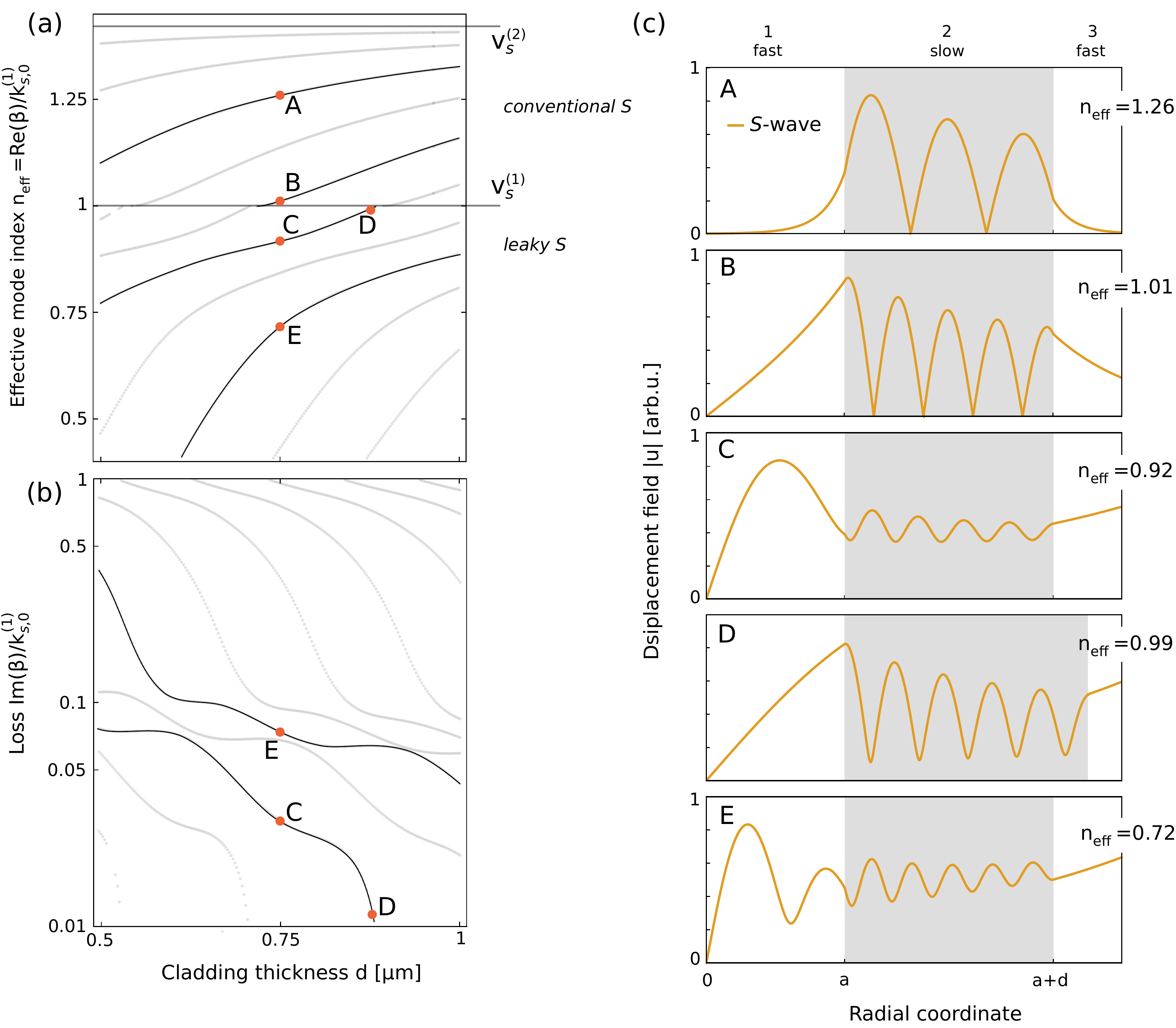}
	\caption{(a) Dispersion, (b) loss, and (c) field distribution plots of azimuthally symmetric ($m=0$) torsional modes of a cylindrical waveguide. Radius of the silicon core is set to 0.5~$\mu$m, and the cladding thickness $d$ (cladding region is marked with gray background in (c)) is normalized by the wavelength of \textit{S}-waves in the cladding material (silicon) $\lambda_s^{(2)}\approx 0.25~\mu$m at the angular frequency of $2\pi\times15$~GHz. Field distributions represent conventionally guided \textit{S} (A,B), and leaky (C-E) modes.}	\label{Fig4}
\end{figure*}

In figure~\ref{Fig4}(a) we present the dispersion of torsional modes, and identify two families of modes: \textit{conventional S}, with $1<\neff<v_\text{s}^{(1)}/v_\text{s}^{(2)}$, and \textit{leaky S}, with $\neff<1$. The azimuthal (and only) component of the displacement field is shown, for a collection of modes, in figure~\ref{Fig4}(c). For A and B, the conventionally guided \textit{S} modes are localized to the cladding layer, and decay exponentially in the outermost layer due to the purely imaginary transverse wavenumber $k_s^{(3)}=\rmi\kappa_s^{(3)}$ (with $\kappa_s^{(3)}>0$). For the leaky modes C-E, we find that the effective mode index \neff~of all the modes increases with the cladding thickness $d$, until the dispersion crosses into the \textit{conventional S} guidance regime. In particular, by tracing the evolution of the branch with modes C and D, we see that as the dispersion approaches the $v_\text{s}^{(1)}$ sound line, oscillations in the cladding layer become more pronounced, and the anti-resonant response becomes significantly stronger, leading to substantial quenching of losses, as observed previously in optical ARROW systems. 

\subsection{Dilatational modes}

\begin{figure*}[htbp!]
	\includegraphics[width=.9\textwidth]{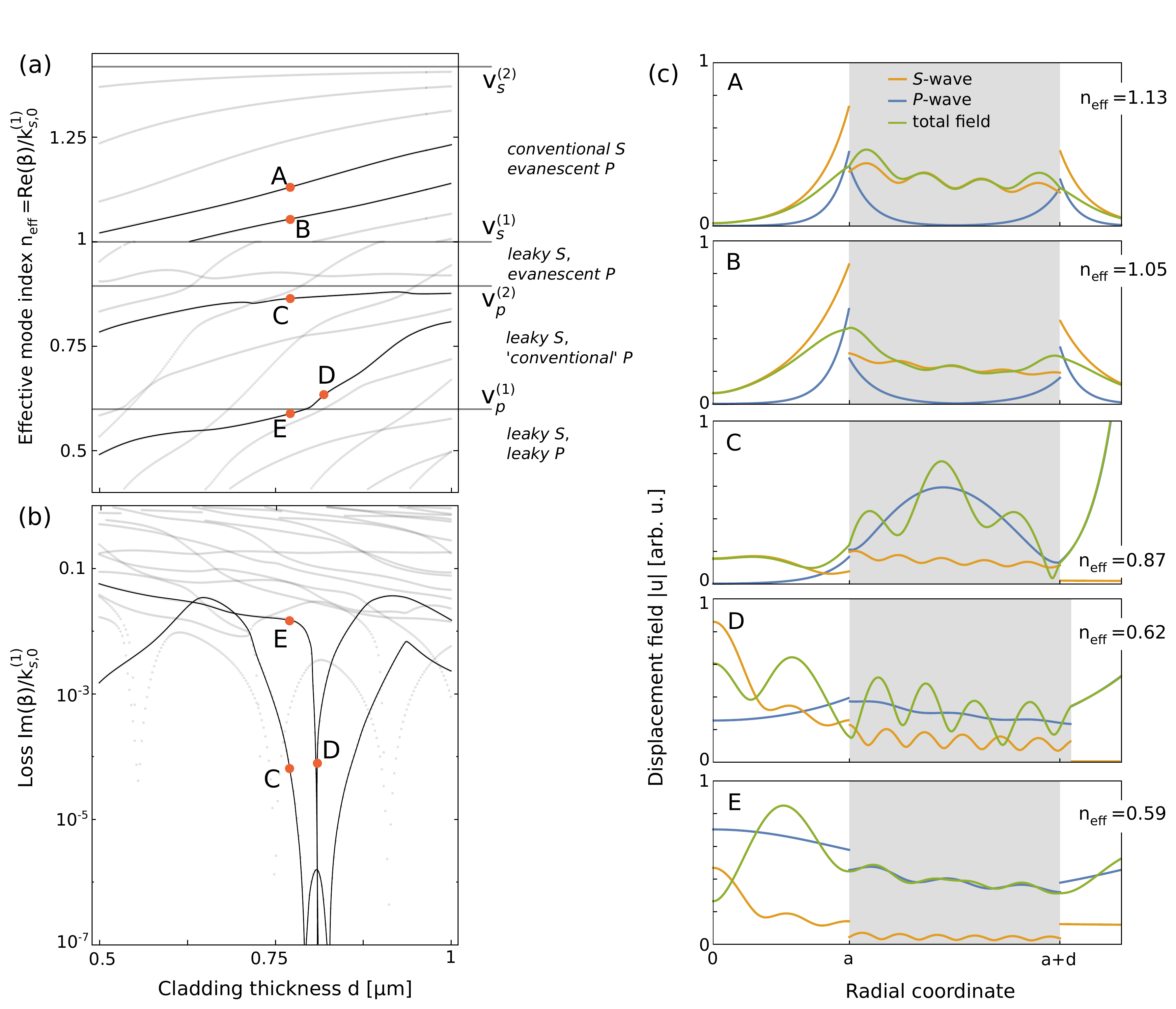}
	\caption{(a) Dispersion, (b) loss and (c) field distribution plots of azimuthally symmetric ($m=0$) dilatational modes of a layered cylindrical waveguide (structure and parameters are identical as in figure~\ref{Fig4}).}
	\label{Fig5}
\end{figure*}

Dilatational modes share the fundamental dispersion characteristics of in-plane modes in planar waveguides, with the four regimes marked in figure~\ref{Fig5}(a) and illustrated in figure~\ref{Fig5}(c), mixing evanescent, conventional, and leaky characteristics of the \textit{S} (orange lines) and \textit{P} (blue lines) waves of the displacement field. \revision{Parameters of the waveguides considered in this calculation are identical to those discussed in section \ref{section31}.}

In particular, for $v_\text{s}^{(1)}/v_\text{p}^{(1)}<\neff<v_\text{s}^{(1)}/v_\text{p}^{(2)}$ we again find the peculiar modes (C and D) characterized by very small loss, for which the \textit{P} waves have the dual characteristic of conventional-like localization to the inner cladding, and leaky-like exponential growth in the outer cladding. For these modes, \textit{S} waves are clearly localized through the anti-resonance to the core of the waveguide.

Furthermore, as for the in-plane modes of the planar waveguide, we expect to find the ARRAW modes right below the fastest sound velocity $v_\text{p}^{(1)}$ line, in the leaky \textit{S} and \textit{P} regime. While these modes are not clearly defined by a dip in loss, we can nevertheless identify ARRAW behavior in the field distributions of the mode. For example, mode E carries the characteristics of both the \textit{S} and \textit{P} components localized to the core, and oscillatory behavior of the components in the cladding layer. 


The above-identified ARRAW modes would clearly not constitute very good acoustic waveguiding channels, as their normalized loss barely reaches $10^{-2}$ (see point F in figure~\ref{Fig3} and point E in figure~\ref{Fig5}). This is mostly due to the fact that the core layers of both the planar and cylindrical systems investigated to this point are too narrow to fit multiple wavelengths of the longer, \textit{P} acoustic waves ($\kpcore a, \kpcore c \ll \pi$). While their geometric parameters were chosen to provide insights into the different regimes of operation of layered acoustic waveguides, we can now consider systems with larger core thicknesses that will provide much better examples of ARRAW guidance, characterized by lower losses and stronger localization of the field to the core, for application in nonlinear Brillouin scattering.

\section{Stimulated Brillouin Scattering in cylindrical ARRAWs}

In the previous section, we considered the acoustic response of cylindrical silicon/silica ARRAWs. Below, we show that such systems, slightly modified to suppress acoustic losses, can simultaneously support propagation of conventionally guided optical waves, and --- thanks to the co-localization of optical and acoustic excitations --- enable nonlinear Brillouin scattering of light propagating through the waveguide. We will focus on the particular case of \textit{Backward Stimulated Brillouin Scattering} (BSBS) \cite{Safavi-Naeini:19,wolff2015stimulated}. In BSBS, two counter-propagating optical modes with wavenumbers $k_i$ and frequencies $\omega_i$ ($i=1,2$) couple via scattering with an acoustic wave characterized by $\beta$ and $\Omega$. The general phase- and frequency-matching conditions
\begin{equation}\label{phase_matching}
	k_1+\mathrm{Re}(\beta) = k_2,~\quad \omega_1 + \Omega = \omega_2,
\end{equation}
can be simplified if we consider the special case of intra-modal BSBS \cite{sipe2016hamiltonian}, in which the counter-propagating optical fields occupy the same mode. Furthermore, since the acoustic frequencies (up to tens of GHz) are much smaller than the \revision{ones corresponding to optical waves in the visible and IR range ($\omega_1\approx \omega_2 \sim 2\pi \times 3\times 10^{14}~\text{Hz}\gg\Omega$)}, we find a simple relationship between the acoustic and optical wavenumbers: $\beta\approx 2|k_1|$.

The nonlinear Brillouin interaction can be used to transfer energy between the two propagating optical fields, at the rate determined by the Brillouin gain $\Gamma$ \cite{Wolff:14,wolff2015stimulated,eggleton2019brillouin,rakichPRX}. In a typical realization, Brillouin interaction amplifies the flux of energy of a weak Stokes optical field $\mathcal{P}^{(S)}$ co-propagating (for Forward SBS) or counter-propagating (BSBS) with respect to a much stronger pump field $\mathcal{P}^{(p)}$ according to $\mathcal{P}^{(S)}(z) = \mathcal{P}^{(S)}(0) \exp\left(\Gamma \mathcal{P}^{(p)} z\right)$. The Brillouin gain coefficient can be expressed as
\begin{equation}\label{Bgain2_maintext}
\Gamma = 4\omega_{1} \frac{Q_m \left|\QPEa+\QMB\right|^2}{\mathcal{E}_b \mathcal{P}^{(1)}\mathcal{P}^{(2)}},
\end{equation}
where $\mathcal{E}_b$ denotes the energy density of the acoustic wave, and $\mathcal{P}^{(i)}$ describes the energy flux of optical mode $i$. \QPEa and \QMB describe two physical processes governing the Brillouin interaction: the photoelastic effect and radiation pressure (referred to as the Moving Boundary, or MB effect), respectively \cite{florez2016brillouin}, localized in the bulk and at the {boundary} of the waveguide. We expect, and verify, that the ARRAWs will primarily enhance the former effect, quantified by the transverse overlap integral between the electric ($\mathbf{e}^{(i)}$) and acoustic ($\mathbf{u}$) fields:
\begin{equation}\label{Q1}
\QPEa = -\varepsilon_0 \int \text{d}^2\mathbf{r}~\varepsilon_a^2 \sum_{ijkl}  [{e}_i^{(1)}]^*{e}_j^{(2)} p_{ijkl} \partial_k u_l^* =\int_0^{\infty} \text{d}r \QPEatilde(r),
\end{equation}
and the Pockels tensor ($p_{ijkl}$). The interaction due to the radiation pressure is typically negligible for waveguides with transverse sizes over 1~$\mu$m \cite{rakichPRX}. Finally, $Q_m$ is the mechanical quality factor, typically used in lieu of the propagation loss, and here defined by the real and imaginary components of the longitudinal acoustic wavenumber $Q_m=\mathrm{Re}(\beta)/[2\mathrm{Im}(\beta)]$.


For simplicity, we focus on the interaction between conventionally guided optical TM, TE and hybrid (HE/EH) modes localized to the core of the waveguide \cite{snyder2012optical}. 


\begin{figure*}[htbp!]
	\includegraphics[width=.9\textwidth]{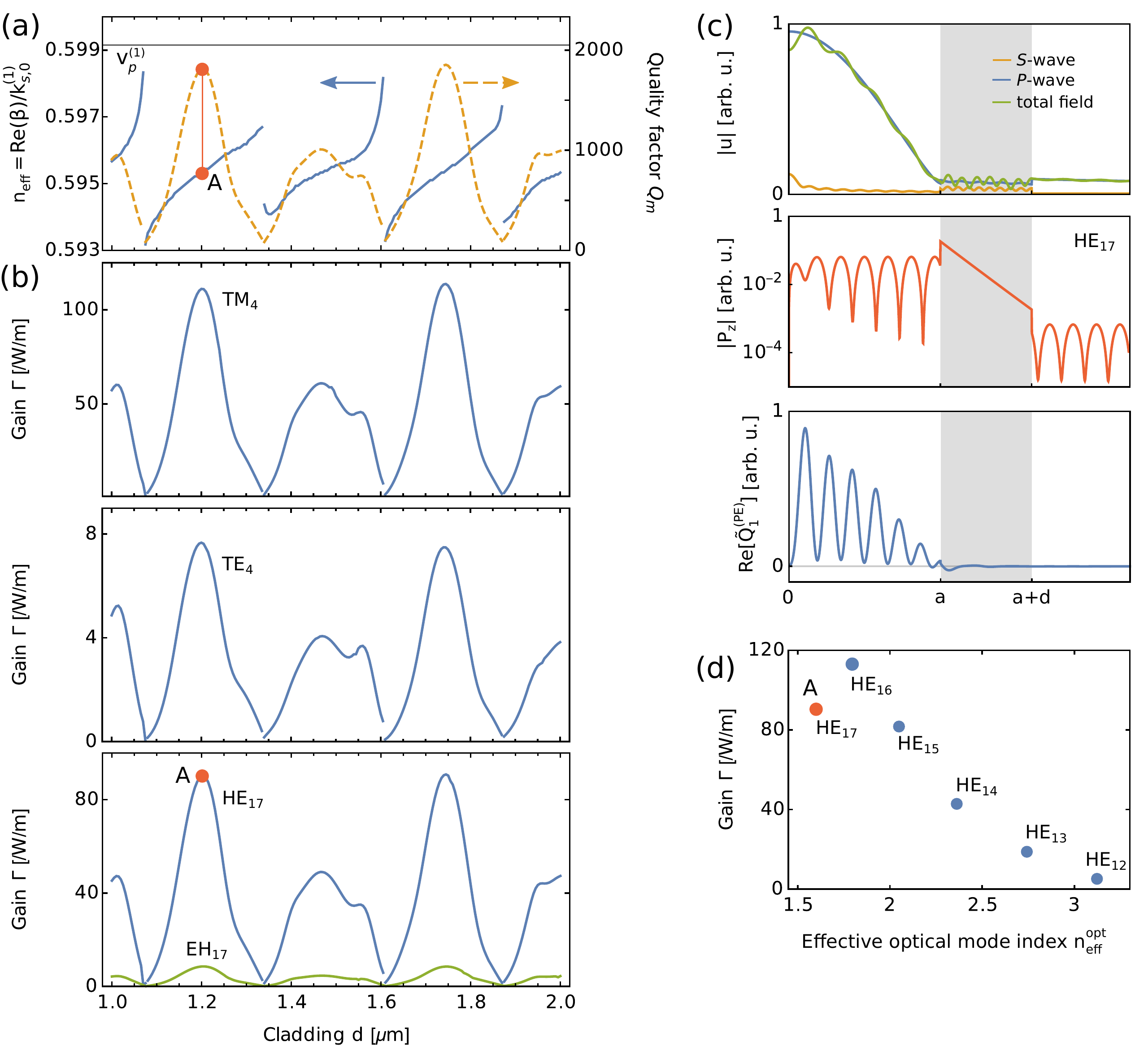}
	\caption{Brillouin interaction between conventionally guided optical and acoustic ARRAW modes of a layered cylindrical waveguide. (a) Dispersion relation (blue lines) and mechanical quality factors $Q_m$ (dashed orange lines) of the highest-$Q_m$ ARRAW modes for increasing cladding thickness $d$ for the cylindrical structure with core radius $a=2~\mu$m and operating at frequency $\Omega/2\pi= 15~$GHz. (b) SBS gain $\Gamma$ calculated for the ARRAW modes identified in (a) mediating interaction between optical TM modes (top panel), TE (central panel) and hybrid HE and EH modes (bottom panel) phase-matched to the acoustic mode \eqref{phase_matching} and characterized by the lowest effective optical mode index $\neff^{\text{opt}}$. (c) SBS interaction between the acoustic mode identified as corresponding to the maximum of $Q_m$ in (a) as A and the HE$_{17}$ optical mode. The three panels represent the displacement field separated into \textit{S} and \textit{P} components, the axial Poynting vector $P_z$ of the optical mode, and the dominant, real part of the overlap function $ \QPEatilde(r) $ originating from the PE interaction defined in \eqref{Q1}. (d) SBS gain between the acoustic mode marked as A in (a) and the phase-matched HE modes of decreasing order (and increasing effective optical mode index) HE$_{17}$ through HE$_{12}$.}
	\label{Fig6}
\end{figure*}

To optimize the BSBS gain, we look for parameters of the system that will simultaneously maximize the mechanical quality factors $Q_m$, and the overlap term \QPEa:

\subsection{Optimizing the mechanical quality factor}
The mechanical quality factor $Q_m$ is optimized by tuning the geometric parameters of the structure (core radius $a$ and cladding thickness $d$), and mechanical frequency $\Omega$. Since the underlying physics of acoustic systems is linear, the modes are invariant under simultaneous rescaling of the geometric parameters and wavelength (or inverse frequency), and consequently, we can consider two of these three parameters as independent.  Here, we fix $\Omega / 2\pi= 15$ GHz, and optimize $a$ and $d$. For clarity, in figure~\ref{Fig6}(a) we show the dispersion relation and mechanical quality factors of the highest-$Q_m$ ARRAW modes lying right below the last sound line ($\neff \leq v_\text{s}^{(1)}/v_\text{p}^{(1)}$), as function of the cladding thickness $d$. \revision{For each of the ARRAW modes denoted with solid blue lines, we identify a resonant dependence of the $Q_m$ on $d$. These features clearly point to the resonances in the reflection within the cladding layer, and the ARRAW nature of the guidance.}

Since we aim to optimize the waveguiding properties of the structure, we consider larger core radii $a=2~\mu$m, which offer mechanical quality factors over 1000 (structures with different core radii are analyzed in Table~\ref{Table1}). As the maximum $Q_m$ increases with the core radius (see results in Table~\ref{Table1}), it is tempting to favour the larger core geometries. However, we should note that our calculations do not account for the viscous losses in the material, which typically limit the $Q_m$ to the order of $\sim 1000$, and thus suppress the advantage of eliminating the acoustic radiative dissipation channels. Furthermore, it was shown in previous reports that the PE contribution to SBS gain exhibits an approximate $a^{-2}$ scaling with the transverse dimension of the waveguide \cite{rakichPRX}. 
\subsection{Overlap term \QPEa}
The particular choice of the optical mode which can undergo BSBS coupling via a selected ARRAW mode is predominantly limited by the phase-matching condition \eqref{phase_matching}, which dictates the magnitude of the longitudinal wavenumber $k_1$ of the optical mode. Therefore, the only parameters which we can optimize here are the type of optical mode (TE, TM or hybrid), and its effective mode index $\neff^{\text{opt}}$. We first focus on the lowest $\neff^{\text{opt}}$ (and high-order) modes supported by the core, and later (figure~\ref{Fig6}(d)) explore the dependence on the order of the optical mode. In particular, in figure~\ref{Fig6}(b) we analyze the BSBS gain coupling between these three types of modes (TM$_4$, TE$_4$ and EH$_{17}$/HE$_{17}$) as a function of cladding thickness --- a parameter which should have little effect on the optical guidance, but governs the ARRAW behavior. We find that the gain closely follows the dependence of the quality factor $Q_m$, suggesting that neither the overlap integral \QPEa nor the normalization factors $\mathcal{E}_b$ and  $\mathcal{P}^{(i)}$ change significantly with $d$.\\

\fulltable{\label{Table1}Backwards SBS gain between acoustic ARRAW modes and optical modes of a silicon/silica/silicon cylindrical waveguide shown in figure~\ref{Fig1}(b). For each waveguide geometry and acoustic frequency $\Omega$ we choose the mode with maximum mechanical quality factor $Q_m$ and calculate BSBS gain coefficient $\Gamma$ (given in \Wm) governing Brillouin coupling between TM$_l$, TE$_l$, HE$_{1l}$ and EH$_{1l}$ optical modes. Optical wavelengths $\lambda$ are given for reference, and setups for which the gain is smaller than 5~\Wm are dismissed.}
\br
geometry & \centre{2}{acoustics} & \centre{12}{optics}  \\ 
\crule{1} & \crule{2} & \crule{12} \\
&& & \centre{3}{TM$_l$} & \centre{3}{TE$_l$} & \centre{3}{HE$_{1l}$}& \centre{3}{EH$_{1l}$} \\
$a/\revision{d}$ & $\Omega/2\pi$ & $Q_m$ & $l$ & $\lambda$ & $\Gamma$ & $l$ & $\lambda$ & $\Gamma$ & $l$ & $\lambda$ & $\Gamma$ & $l$ & $\lambda$ & $\Gamma$ \\
($\mu$m) & (GHz) &&  & ($\mu$m) & &  & ($\mu$m)& &  & ($\mu$m)&  &  & ($\mu$m)&  \\
\mr
1.5/2.82 & 15  & 570         & 4 & 1.89 & 23.5    & 4--0 & -- & $<5$    & 7 & 2.03 & 42.6     & 7--1 & -- & $<5$ \\
    &    &                   & 3 & 2.19 & 44.3    &  &  &             & 6 & 2.39 & 37.8    &  &   &  \\ 
    &    &                   & 2 & 2.62 & 28.1    &  &  &             & 5  & 3.50 & 5.2               &  &   &   \\  
    &    &                   & 1 & 3.17 & 10.8    &  &  &             & 4--2 & -- & $<5$               &  &   &   \\
    &    &                   & 0 & 3.80 & $<5$    &  &  &             &   &   &               &  &   &   \\\mr
2/1.2 & 15 & 1815 & 6 & 1.83 & 46.5 & 6 & 1.85 & 7.6    & 7 & 1.92 & 90.2   & 7 & 1.96 & 8.5 \\ 
  &    &      & 5 & 2.02 & 110.9     & 5 & 2.08 & 6.6    & 6 & 2.15 & 113.4  & 6 & 2.22 & 6.5 \\ 
  &    &      & 4 & 2.29 & 99.1      & 4--0 & -- & $<5$    & 5 & 2.45 & 81.9   & 5--1 & -- & $<5$  \\
  &    &      & 3 & 2.63 & 62.8      &  &   &            & 4 & 2.83 & 43.3   &  &   &   \\
  &    &      & 2 & 3.04 & 30.8      &   &   &           & 3 & 3.27 & 17.9   &  &   &   \\
  &    &      & 1 & 3.50 & 11.6     &   &   &           & 2 & 3.74 & 5.4    &  &   &   \\
  &    &      & 0 & 3.94 & $<5$     &   &   &           & 1 & -- & $<5$    &  &   &   \\ \mr
2.5/1.74 & 15 & 4247 & 7 & 1.94 & 177.9   & 7 & 1.97 & 13.0    & 9 & 1.86 & 121.2    & 9 & 1.88 & 15.0 \\ 
    &    &           & 6 & 2.13 & 195.8   & 6 & 2.18 & 10.5    & 8 & 2.03 & 200.5    & 8 & 2.07 & 12.5 \\ 
    &    &           & 5 & 2.36 & 151.2   & 5 & 2.42 & 7.9     & 7 & 2.24 & 175.5    & 7 & 2.30 & 9.6 \\
    &    &           & 4 & 2.64 & 98.7    & 4--0 & -- & $<5$     & 6 & 2.49 & 122.6    & 6 & 2.87 & 5.1 \\
    &    &           & 3 & 2.96 & 55.6    &  &  &              & 5 & 2.79 & 73.1     & 5--1 & -- & $<5$ \\
    &    &           & 2 & 3.32 & 26.8    &  &  &              & 4 & 3.14 & 37.5     &  &  &  \\
    &    &           & 1 & 3.70 & 10.8   &  &  &               & 3 & 3.52 & 15.9     &  &  &  \\
    &    &           & 0 & 4.01 & $<5$   &  & &                & 2 & 3.87 & 5.6      &  &  &  \\
    &    &           &  &  &    &  & &                & 1 & -- & $<5$        &  &  &  \\ \mr
1/1.14 & 30 & 1815 & 6 & 0.91 & 294.7    & 6 & 0.92 & 58.7    & 7 & 0.96 & 704.8       & 7 & 0.98 & 65.2 \\ 
  &  &             & 5 & 1.01 & 891.4    & 5 & 1.04 & 54.8    & 6 & 1.07 & 888.3       & 6 & 1.11 & 49.2 \\ 
  &  &             & 4 & 1.15 & 793.4    & 4 & 1.18 & 40.0    & 5 & 1.23 & 628.6       & 5 & 1.27 & 34.3 \\
  &   &            & 3 & 1.31 & 492.8    & 3 & 1.36 & 27.2    & 4 & 1.41 & 338.7       & 4 & 1.46 & 23.0  \\
  &   &            & 2 & 1.52 & 240.8    & 2 & 1.56 & 18.0    & 3 & 1.64 & 140.2       & 3 & 1.68  & 15.1  \\
  &   &            & 1 & 1.75 & 93.5     & 1 & 1.78 & 12.0    & 2 & 1.87 & 42.1        & 2 & 1.91 & 8.9  \\
  &   &            & 0 & 1.97 & 25.2     & 0 & 1.98 & 8.1     & 1 & 2.04 & 16.6        & 1 & -- &  $<5$ \\ \br
\endfulltable

The orange dot in the plots in figure~\ref{Fig6}(a,b) denotes the parameters of optical and acoustic modes analyzed in detail in figure~\ref{Fig6}(c). The radial profile of the acoustic field, shown in the top panel of figure~\ref{Fig6}(a), indicates significant localization to the core --- as expected for large-$Q_m$ ARRAW modes. Together with the core-localized optical mode (middle panel), this results in a strong localization of the \QPEatilde(r) overlap integral kernel shown in the bottom panel.

To further enhance the Brillouin gain, we can consider changing the geometric parameters of the waveguide (e.g. core radius $a$), operating mechanical frequency $\Omega$, or explore coupling to different orders of the optical modes. We provide a comparison of selected BSBS ARRAW systems in Table~\ref{Table1}, and find a few guiding principles for designing such systems:
\begin{itemize}
    \item as reported by Rakich \etal \cite{rakichPRX}, the smaller cross section waveguides yield larger BSBS gain --- however, this principle trades off against the sharp decrease in mechanical quality factor for small core radii; simultaneous increase of mechanical frequencies (e.g. towards $30~$GHz frequency) should allow us to retain high $Q_m$'s due to the linear nature of the acoustic physics, but the overall gain would likely become suppressed by the increased non-radiative acoustic losses at higher frequencies,
    \item dependence on the order of optical mode (or effective optical mode index) is not monotonic, and in fact is the smallest for the most homogeneous field of the lowest order mode; this dependence is shown, for a number of hybrid modes from HE$_{17}$ to HE$_{12}$, in figure~\ref{Fig6}(d).
\end{itemize}



\subsection{Comparison to other BSBS waveguides}

It is instructive to compare the nonlinear performance of the investigated ARRAW to other BSBS waveguiding systems. The mechanical quality factor $Q_m$ of our structure can easily reach $10^3$, which is a result comparable to that found in suspended waveguides, in which the radiative dissipation of acoustic waves is suppressed, and $Q_m$ becomes limited by the intrinsic viscous losses in silicon \cite{schmidt19,van2015interaction}. Furthermore, the maximum Brillouin gain found in our system (nearly 1000~\Wm) is of a similar order to that reported in GHz BSBS systems, including those based on sub-micron silicon slot waveguides where the optoacoustic interaction is dominated by radiation pressure \cite{doi:10.1063/1.4955002}, or relying on materials with different optoacoustic properties, such as chalcogenides \cite{Pant:11}.

\subsection{Outlook}
Our simple designs can be further modified to better suit integrated platforms by considering finite outer-cladding layers or multiple anti-resonant layers. The principles of ARRAW guidance can also be combined with other mechanisms, for example by using the anti-resonant reflection to suppress the acoustic dissipation from exposed cores of rib waveguides into the substrate. Alternatively, the entire designs could be reversed and implement optical anti-resonant guidance and conventional acoustic TIR.

\section{Summary}

We have proposed a new type of multilayered optoacoustic Anti-Resonant Reflecting Acoustic Waveguide capable of supporting the simultaneous and co-localized guidance of GHz acoustic and near-IR optical signals. While the optical waves are TIR-guided in the high-refractive index core, the acoustic waves are localized to the core by anti-resonant reflection in the inner cladding layer. This mechanism can be harnessed to efficiently suppress the dissipation of acoustic waves into the outermost layers, and enable efficient Brillouin scattering between the counter-propagating optical waves. Our estimates indicate that silicon/silica ARRAWs can match the record performance of backward stimulated Brillouin scattering in silicon/silica platforms without relying on sub-micron confinement of fields or interactions induced by radiation pressure.



\ack
Authors acknowledge funding from Australian Research Council (ARC) (Discovery Project DP160101691) and the Macquarie University Research Fellowship Scheme (MQRF0001036). 

\appendix
\setcounter{section}{0}

\section{Planar waveguide}
\label{AppendixA}

In this section we present the transfer matrix method used to derive transcendental conditions for waveguiding modes of the planar structure shown in figure~\ref{Fig1}(a).

\subsection{Out-of-plane polarization}


We start with the simpler case of out-of-plane polarization, in which modes are composed of shear waves only, and displacement fields have $u_y$ components only, meaning that the boundary conditions demand the continuity of $u_y$ components of the displacement field and continuity of the $T_{xy}$ element of the stress tensor.

We use the ansatz
\begin{equation}
\mathbf{u}^{(j)}(x,z,t) = \hat{\mathbf{y}}\left({u}_+^{(j)} \rme^{\rmi\ksj x} + {u}_-^{(j)} \rme^{-\rmi\ksj x}\right)\rme^{\rmi (\beta z-\Omega t)} + \text{c.c.},
\end{equation}
where ${u}_+^{(j)} $ and ${u}_-^{(j)}$ are undetermined coefficients for the right- and left-propagating fields in the $j$\textsuperscript{th} layer respectively.
We can thus represent the boundary conditions at $x=x_j$ between the $j$th and $(j+1)$th layer as
\begin{align}\label{M.defs}
&\underbrace{\begin{bmatrix}
	\rme^{\rmi\ksj x_j} && \rme^{-\rmi\ksj x_j} \\
	\ksj\mu^{(j)} \rme^{\rmi\ksj x_j} && -\ksj\mu^{(j)} \rme^{-\rmi\ksj x_j}
	\end{bmatrix}}_{M^{(j)}(x_j)}  
\underbrace{\begin{bmatrix}
	{u}_+^{(j)} \\
	{u}_-^{(j)}
	\end{bmatrix}}_{\tilde{\mathbf{u}}^{(j)}} \\ \nonumber
&= 
\underbrace{\begin{bmatrix}
	\rme^{\rmi k_s^{(j+1)} x_j} && \rme^{-\rmi k_s^{(j+1)} x_j} \\
	k_s^{(j+1)}\mu^{(j+1)} \rme^{\rmi k_s^{(j+1)} x_j} && -k_s^{(j+1)}\mu^{(j+1)} \rme^{-\rmi k_s^{(j+1)} x_j}
	\end{bmatrix}}_{M^{(j+1)}(x_j)} 
\underbrace{\begin{bmatrix}
	{u}_+^{(j+1)} \\
	{u}_-^{(j+1)}
	\end{bmatrix}}_{\tilde{\mathbf{u}}^{(j+1)}}.
\end{align}
We now multiply both sides by $\left[M^{(j+1)}(x_j)\right]^{-1}$ to obtain
\begin{equation}\label{def.S}
\left[M^{(j+1)}(x_j)\right]^{-1}M^{(j)}(x_j)\tilde{\mathbf{u}}^{(j)} \doteq S^{(j)}\tilde{\mathbf{u}}^{(j)}  = \tilde{\mathbf{u}}^{(j+1)}.
\end{equation}
This formalism relates the fields in the last medium $j=5$ with the first medium $j=1$ ($S^{(\text{all})}\tilde{\mathbf{u}}^{(1)} =\tilde{\mathbf{u}}^{(5)}$), through the matrix
\begin{equation}\label{construct.S}
S^{(\text{all})} = S^{(4)} S^{(3)} S^{(2)} S^{(1)} .
\end{equation}

To find waveguiding modes, we require that fields in the first and the last mediums should be \textit{outgoing} only, i.e.
\begin{equation}
\tilde{\mathbf{u}}^{(1)} =\begin{bmatrix}
0 \\ {u}_-^{(1)}
\end{bmatrix}, \quad 
\tilde{\mathbf{u}}^{(5)} =\begin{bmatrix}
{u}_+^{(5)} \\ 0
\end{bmatrix}.
\end{equation}
This condition only holds when $S^{(\text{all})}_{22} = 0$, which defines a  4th order polynomial in the quantities $k_s^{(i)}$. We can simplify it by considering the symmetric and anti-symmetric modes of the structure separately. Expressing the field $\mathbf{u}^{(3)}$ through the transfer matrix, imposing the above symmetry conditions, and accounting for the vanishing of the incident wave in medium 1, we arrive at the resonance condition:
\begin{equation}
    \left( W^{\pm} S^{(2)} S^{(1)} \right)_{22} = 0, \label{eq:symm_cond_inplane}
\end{equation}
where $W^{\pm}$ imposes the symmetry/antisymmetry of $\mathbf{u}^{(3)}$ at $x=0$, and is defined by
\begin{equation}
    W^{\pm} = \begin{bmatrix}
    1 && 0 \\
    1 && \pm 1
    \end{bmatrix}.
\end{equation}
The condition \eqref{eq:symm_cond_inplane} is equivalent to equations \eqref{sym} and \eqref{asym}. 
\subsection{In-plane polarization}
For in-plane polarization, the displacement fields lie in the $\hat{\mathbf{x}}\hat{\mathbf{y}}$ plane, and they have both longitudinal and transverse contributions. Importantly, these two contributions exhibit different transverse wavevectors. The displacement field components thus have two right- and left-propagating field coefficients each ($u_{\text{p},\pm}^{(j)}$ and $u_{\text{s},\pm}^{(j)}$):
\begin{align}
\mathbf{u}^{(j)}_\text{p}(x,z,t) &= \left[\left(\hat{\mathbf{x}}\klj+\hat{\mathbf{z}}\beta\right) {u}_{\text{p}+}^{(j)} \rme^{\rmi \klj x} \right.\\ \nonumber  &\qquad + \left. \left(-\hat{\mathbf{x}}\klj+\hat{\mathbf{z}}\beta\right) {u}_{\text{p}-}^{(j)} \rme^{-\rmi \klj x}\right]\rme^{\rmi (\beta z-\Omega t)} + \text{c.c.}, \\
\mathbf{u}^{(j)}_\text{s}(x,z,t) &= \left[\left(-\hat{\mathbf{x}}\beta+\hat{\mathbf{z}}k^{(j)}_s\right) {u}_{\text{s}+}^{(j)} \rme^{\rmi \ksj x} \right.\\ \nonumber  &\qquad + \left. \left(\hat{\mathbf{x}}\beta+\hat{\mathbf{z}}\ksj\right) {u}_{\text{s}-}^{(j)} \rme^{-\rmi \ksj x}\right]\rme^{\rmi (\beta z-\Omega t)}+ \text{c.c.}, \\
\mathbf{u}^{(j)}(x,z,t) &= \mathbf{u}^{(j)}_\text{p}(x,z,t) + \mathbf{u}^{(j)}_\text{s}(x,z,t).
\end{align}
The relevant elements of the stress tensor which need to be continuous at each interface are $T_{xx}^{(j)}$, and $T_{xz}^{(j)}$.

The matrices $M^{(j)}(x)$ and vectors $\tilde{\mathbf{u}}^{(j)}$ are thus given by:
\begin{equation}
\tilde{\mathbf{u}}^{(j)} = \begin{bmatrix}
{u}_{\text{p}+}^{(j)} && {u}_{\text{p}-}^{(j)} && {u}_{\text{s}+}^{(j)} && {u}_{\text{s}-}^{(j)}
\end{bmatrix}^T,
\end{equation}
\begin{equation}
M^{(j)}(x) = 
\begin{bmatrix}
\klj \rme^{\rmi\klj x} && -\klj \rme^{-\rmi \klj x} && -\beta \rme^{\rmi\ksj x} && \beta \rme^{-\rmi \ksj x} \\
\beta \rme^{\rmi\klj x} && \beta \rme^{-\rmi \klj x} && \ksj \rme^{\rmi\ksj x} && \ksj \rme^{-\rmi \ksj x} \\
m_{31} && m_{32} && m_{33} && m_{34} \\
m_{41} && m_{42} && m_{43} && m_{44}
\end{bmatrix},
\end{equation}
with 
\begin{eqnarray}
m_{31} &= \left[(\lambda^{(j)}+2\mu^{(j)})(\klj)^2 + \lambda^{(j)} \beta^2\right] \rme^{\rmi \klj x},\\
m_{32} &= \left[(\lambda^{(j)}+2\mu^{(j)})(\klj)^2 + \lambda^{(j)} \beta^2\right] \rme^{-\rmi  \klj x},\\
m_{33} &= -2\mu^{(j)}\beta \ksj \rme^{\rmi \ksj x},\\
m_{34} &= -2\mu^{(j)}\beta \ksj \rme^{-\rmi  \ksj x},\\
m_{41} &= 2 \mu^{(j)} \beta \klj \rme^{\rmi \klj x},\\
m_{42} &= -2 \mu^{(j)} \beta \klj \rme^{-\rmi  \klj x},
\\
m_{43} &= \mu^{(j)}  \left[\left(\ksj\right)^2-\beta^2\right] \rme^{\rmi \ksj x},\\
m_{44} &= -\mu^{(j)}  \left[\left(\ksj\right)^2-\beta^2\right] \rme^{-\rmi  \ksj x}.
\end{eqnarray}
As in the previous section, this formulation of matrices $M^{(j)}$ allows us to construct matrices $S^{(j)}$ as in \eqref{construct.S}, which will readily mix the \textit{S} and \textit{P} components of the fields in layers $j$ and $j+1$. This behavior signals coupling between the \textit{S} and \textit{P} waves through scattering at interfaces.

To find waveguiding modes, we again require that fields in the first and the last layers should be \textit{outgoing} only, i.e.
\begin{equation}
\tilde{\mathbf{u}}^{(5)} =\begin{bmatrix}
{u}_{\text{p}+}^{(5)} \\ 0 \\ {u}_{\text{s}+}^{(5)} \\ 0
\end{bmatrix}, \quad \tilde{\mathbf{u}}^{(1)} =\begin{bmatrix}
0 \\ {u}_{\text{p}-}^{(1)} \\ 0 \\ {u}_{\text{s}-}^{(1)}
\end{bmatrix}.
\end{equation}
Thus, we formulate the transfer problem as $\tilde{\mathbf{u}}^{(5)} =  S^{(\text{all})}\tilde{\mathbf{u}}^{(1)},$ and look for the solutions to
\begin{equation}
\det \begin{bmatrix}
(S^{(\text{all})})_{22} & (S^{(\text{all})})_{24}\\
(S^{(\text{all})})_{42} & (S^{(\text{all})})_{44}\\
\end{bmatrix} = 0, \label{eq:det_eq_transc}
\end{equation}
for discrete values of $\beta$ and fixed $\Omega$. As discussed in the main text, this transcendental equation will have both real and complex solutions, corresponding to conventional and ARRAW modes, respectively. Treating the real and imaginary parts of $\beta$ as independent variables and applying one of the many standard multi-dimensional root-finding algorithms is computationally intensive and time-consuming. However, by treating $\beta$ as a complex variable, we are able to apply methods of complex analysis; \eqref{eq:det_eq_transc} is analytic everywhere except along a branch cut on the real axis. In order to calculate the dispersion relations presented in the main text, we applied the \textit{Global Complex Roots and Pole Finding} (GRPF) algorithm developed in \cite{grpf}.
%
\section{Cylindrical waveguide}
\label{AppendixB}

In this section we present the transfer matrix method used to derive transcendental conditions for waveguiding modes of the cylindrical structures shown in figure~\ref{Fig1}(b).

\subsection{Basis of modes for a layered cylindrical waveguide}
The problem in question is that of a cylindrical rod of isotropic material and infinite length, as shown in figure~\ref{Fig1}(b). The problem is solved using cylindrical coordinates $(r,\theta,z)$ and assuming a time-harmonic solution of the form
\begin{equation}\label{harmonicsolution.cylinder}
	\mathbf{u}(r,\theta,z,t) = \mathbf{U}(r,\theta) \rme^{\rmi(\beta z - \Omega t)} + \text{c.c.}
\end{equation}
The relevant elements of the stress tensor in cylindrical coordinates are identified as \cite{auld1973acoustic}
\begin{align}\label{stresses.cylindrical}
T_{rr} &= \lambda \left( \del_r u_r + \frac{1}{r}\del_\theta u_\theta + \frac{u_r}{r} + \del_z u_z \right) + 2\mu \del_r u_r, \\
T_{r\theta} &= \mu\left(\frac{1}{r}\del_\theta u_r - \frac{u_\theta}{r} + \del_r u_\theta\right), \\
T_{rz} &= \mu \left(\del_z u_r + \del_r u_z\right).\label{eq:stressfield}
\end{align}
A convenient expansion for the modes of the cylinder can be derived by expressing the displacement fields $\mathbf{u}$ through two scalar potentials $\Phi$ and $\Psi$ as 
\begin{equation}\label{general.u.cylindrical}
	\mathbf{u} = A\nabla \Phi + B\nabla\times(\hat{\mathbf{z}}\Psi) + C\nabla\times\nabla\times(\hat{\mathbf{z}}\Psi),
\end{equation}
where $A$, $B$ and $C$ are free coefficients \cite{auld1973acoustic}. These potentials are solutions to the scalar wave equations
\begin{align}
	&\left[ \nabla_\perp^2+\left(\frac{\partial^2}{\partial z^2} - \frac{1}{v_\text{p}^2}\frac{\partial^2}{\partial t^2}\right)\right]\Phi(r,\theta,z,t)=\left[\nabla_\perp^2+\left(\frac{\Omega^2}{v_\text{p}^2}-\beta^2\right)\right]\phi(r,\theta)\rme^{\rmi(\beta z - \Omega t)} = 0,\\
	&\left[\nabla_\perp^2+\left(\frac{\partial^2}{\partial z^2} - \frac{1}{v_\text{s}^2}\frac{\partial^2}{\partial t^2}\right)\right]\Psi(r,\theta,z,t)=\left[\nabla_\perp^2+\left(\frac{\Omega^2}{v_\text{s}^2}-\beta^2\right)\right]\psi(r,\theta)\rme^{\rmi(\beta z - \Omega t)} = 0,
\end{align}
where we have made a similar ansatz for the scalar fields as for $\mathbf{u}$ (see \eqref{harmonicsolution.cylinder}). The resulting equations for the transverse scalar fields $\phi$ and $\psi$ can be solved by supposing an azimuthal dependence of $\rme^{im\theta}$, which yields the radial dependence of the potentials as solutions to the Bessel equation $\chi_m(k_i r)$ (we will comment on the particular choice of $\chi$), where $k_i^2=(\Omega/v_i)^2-\beta^2$ for $i=\text{p}$ ($\phi$) or $\text{s}$ ($\psi$).

We thus arrive at the general form of the displacement field:
\begin{align}
	&U_r = A k_\text{p} \chi'_m(k_\text{p} r) + B \rmi\beta k_\text{s} \chi'_m(k_\text{s} r) + C \rmi m \frac{\chi_m(k_\text{s} r)}{r}, \label{eq:gensol3}\\
	&U_\theta = A \rmi m \frac{\chi_m(k_\text{p} r)}{r} - B m\beta \frac{\chi_m(k_\text{s} r)}{r} - C k_s \chi_m'(k_\text{s} r), \label{eq:gensol4}\\
	&U_z = A \rmi\beta \chi_m(k_\text{p} r) + B k_\text{s}^2 \chi_m(k_\text{p} r). \label{eq:gensol5}
\end{align}
Therefore, we can observe that in the axially symmetric case, when $m=0$, the displacement field decouples into \textit{torsional} modes with only a $u_\theta$ component ($A=B=0$), and \textit{dilatational} modes ($C=0$) with $u_r$ and $u_z$ components. We will focus on this case throughout.
\subsection{Torsional modes}
Just as for the planar waveguide structure, we can decompose the displacement field in each layer into a basis of outgoing and incoming waves. This is determined by the choice of the functions $\chi$ from the family of Bessel and Hankel functions. To represent the outgoing wave, we choose $\chi$ in the form of Hankel functions of the first kind $H_0^{(1)}$; we will suppress the $(1)$ superscript in all the following equations for clarity. To describe the incoming waves, we could use the Hankel functions of the second kind. However, since $H^{(2)}$ diverge at $r=0$, we instead consider the Bessel functions of the first kind $J^{(1)}$ --- while this component will not strickly describe \textit{incoming} waves only, our expansion will remain complete. The displacement field is thus given by
\begin{equation}
    \mathbf{u}^{(j)}(r,z,t) = \hat{\bm{\theta}}\left({u}_+^{(j)} H'_0(\ksj r) + {u}_-^{(j)} J'_0(\ksj r)\right)\rme^{\rmi (\beta z-\Omega t)}.
\end{equation}
Here the derivatives of special Bessel and Hankel functions (denoted by $'$) are calculated with respect to the entire argument of the respective function. The only relevant non-zero component of the stress tensor is $T_{r\theta}$. Thus, the matrices $M^{(j)}(r)$ and coefficient vectors $\tilde{\mathbf{u}}^{(j)}$ defined in \ref{AppendixA} are given by
\begin{equation}
    \tilde{\mathbf{u}}^{(j)} = \begin{bmatrix}
    {u}_+^{(j)} && {u}_-^{(j)}
    \end{bmatrix}^T,
\end{equation}
\begin{align}
	M^{(j)}(r) &= \begin{bmatrix}
	H_{0}'(\ksj r) &&  J_{0}'(\ksj r) \\
	\mu^{(j)} \ksj (2H_{0}''(\ksj r) + H_{0}(\ksj r)) && \mu^{(j)} \ksj (2J_0''(\ksj r) + J_{0}(\ksj r))
	\end{bmatrix}\\ 
	&= \begin{bmatrix}
	-H_{1}(\ksj r) &&  -J_{1}(\ksj r) \\
	\mu^{(j)} \ksj H_{2}(\ksj r) && \mu^{(j)} \ksj J_2(\ksj r)
	\end{bmatrix}. \nonumber
\end{align}
The waveguiding condition is formulated in exactly the same manner as for the planar structure, by connecting the vectors of coefficients in the outermost $\tilde{\mathbf{u}}^{(3)}$ and innermost layer $\tilde{\mathbf{u}}^{(1)}$ through the $M$ matrices (see \ref{def.S}), and requiring that the innermost field is non-sigular at $r=0$ (no outgoing wave in layer $(1)$; $u_+^{(1)}=0$) and the outermost field has no incoming wave components (since the Bessel function can be expressed as a sum of incoming and outgoing waves, we put $u_-^{(3)}=0$).
\subsection{Dilatational modes}
Following the same arguments as in the previous subsection, we again write down the displacement field as a sum of `+' and `-' components, with each one including contributions from both the \textit{P} and \textit{S} waves:
\begin{align}
    \mathbf{u}^{(j)}_\text{p}(r,z,t) &= \left[\left(\hat{\mathbf{r}} \klj H_{0\text{p}}' + \hat{\mathbf{z}} \rmi\beta H_{0\text{p}} \right) {u}_{p+}^{(j)} \right. \\ \nonumber &\qquad \left.+ \left(\hat{\mathbf{r}} \klj J_{0\text{p}}' + \hat{\mathbf{z}} \rmi\beta J_{0\text{p}} \right) {u}_{p-}^{(j)}\right]\rme^{\rmi (\beta z-\Omega t)} + \text{c.c.}, \nonumber \\
    \mathbf{u}^{(j)}_\text{s}(x,z,t) &= \left[\left(\hat{\mathbf{r}} \rmi \beta H_{0\text{s}}' + \hat{\mathbf{z}} \ksj H_{0\text{s}} \right) {u}_{s+}^{(j)} \right. \\ \nonumber  &\qquad \left.+ \left(\hat{\mathbf{r}} \rmi \beta J_{0\text{s}}' + \hat{\mathbf{z}} \ksj J_{0\text{s}} \right) {u}_{s-}^{(j)}\right]\rme^{\rmi (\beta z-\Omega t)}+ \text{c.c.},\nonumber \\
    \mathbf{u}^{(j)}(x,z,t) &= \mathbf{u}^{(j)}_\text{p}(x,z,t) + \mathbf{u}^{(j)}_\text{s}(x,z,t),
\end{align}
where we have introduced the abbreviations $H_{0i} = H_0 (k_i^{(j)} r)$ and $J_{0i} = J_0 (k_i^{(j)} r)$, for $i=\text{s},\text{p}$. The relevant components of the stress tensor are $T_{rr}$ and $T_{rz}$. Therefore, we obtain the coefficients vectors
\begin{equation}
    \tilde{\mathbf{u}}^{(j)} = \begin{bmatrix}
    {u}_{\text{p}+}^{(j)} && {u}_{\text{p}-}^{(j)} && {u}_{\text{s}+}^{(j)} && {u}_{\text{s}-}^{(j)}
    \end{bmatrix}^T,
\end{equation}
and matrices $M$ for each layer $j$
\begin{equation}
    M^{(j)} (r) = \begin{bmatrix}
    \klj H_{0\text{p}}' && \klj J_{0\text{p}}' && \rmi \beta H_{0\text{s}}' && \rmi \beta J_{0\text{s}}' \\
    \rmi \beta H_{0\text{p}} && \rmi \beta J_{0\text{p}} && \ksj H_{0\text{s}} && \ksj J_{0\text{s}} \\
    m_{31} && m_{32} && m_{33} && m_{34} \\
    m_{41} && m_{42} && m_{43} && m_{44}
\end{bmatrix},
\end{equation}
where
\begin{eqnarray}
    m_{31} &= 2\mu^{(j)} (\klj)^2 H_{0\text{p}}'' - \lambda^{(j)} \left((\klj)^2 + \beta^2 \right)H_{0\text{p}},\\ 
    m_{32} &= 2\mu^{(j)} (\klj)^2 J_{0\text{p}}'' - \lambda^{(j)} \left((\klj)^2 + \beta^2 \right)J_{0\text{p}},\\
    m_{33} &= 2\rmi\beta \mu^{(j)} (\ksj)^2 H_{0\text{s}}'',\\
    m_{34} &= 2\rmi\beta \mu^{(j)} (\ksj)^2 J_{0\text{s}}'',\\ 
    m_{41} &= 2\rmi\beta \mu^{(j)} \klj H_{0\text{p}}',\\
    m_{42} &= 2\rmi\beta \mu^{(j)} \klj J_{0\text{p}}',\\ 
    m_{43} &= \mu^{(j)} \ksj \left((\ksj)^2 - \beta^2 \right) H_{0\text{s}}',\\  
    m_{44} &= \mu^{(j)} \ksj \left((\ksj)^2 - \beta^2 \right)J_{0\text{s}}'.
\end{eqnarray}
Again, we formulate the waveguiding condition by relating vectors $\tilde{\mathbf{u}}^{(1)}$ and $\tilde{\mathbf{u}}^{(3)}$ via $M$ matrices, and require that the innermost (outermost) layer has no outoing (incoming) components. The resulting transcendental equation can also be solved numerically using the GRPF algorithm.

\revision{
\section{Material properties of silicon and silica}
\label{AppendixC}

Throughout this work, we use isotropic models for both silicon and silica, characterized with the properties listed in table~\ref{table.materials}.
}

\fulltable{
\revision{Elastic and photoelastic properties of silicon and silica \label{table.materials}}}
\br
\revision{Material} & \revision{Density} & \revision{Young modulus} & \revision{Poisson ratio} & \revision{Photoelastic tensor}  \\ 
& \revision{(kg/m$^3$)} & \revision{[GPa]} &  & \revision{(p$_{11}$, p$_{12}$, p$_{44}$)}\\ \br
\revision{Silicon} & \revision{2329} & \revision{162} & \revision{0.22} & \revision{(-0.094, 0.017, -0.051)} \\ \mr 
\revision{Silica} & \revision{2203} & \revision{73.1} & \revision{0.17} & \revision{(0.12, 0.27, -0.075)} \\
\br
\endfulltable

\revision{
The isotropic models for silica and silicon are an approximation included to facilitate the development of the analytical formalism presented in this work, based on published data \cite{dolbow1996effect,weber2018handbook}. Future applications and design of optoacoustic waveguides implementing the principles or ARRAW guidance will likely require a full numerical solution based on the anisotropic description of the stiffness tensor of both the materials.
}

\section*{References}

\bibliographystyle{unsrt} 
\bibliography{bibliography_fixed}

\end{document}